\documentclass{ws-procs9x6}
\usepackage{color}
\begin{document}

\title{Observation of Kink Instability as Driver of Recurrent Flares in
AR 10960}

\author{A.K. Srivastava$^*$}
\address{Aryabhatta Research Institute of Observational Sciences (ARIES), Manora Peak,
Nainital-263 129, India.\\
$^*$E-mail: aks@aries.res.in}

\author{Pankaj Kumar}
\address{Aryabhatta Research Institute of Observational Sciences (ARIES), Manora Peak,
Nainital-263 129, India.}

\author{T.V. Zaqarashvili}
\address{Space Research Institute, Austrian Academy of Sciences, Graz 8042, Austria.}
\address{Abastumani Astrophysical Observatory at Ilia State University, Al Kazbegi ave. 2a, 0160 Tbilisi, Georgia.}

\author{B.P. Filippov}
\address{Pushkov Institute of Terrestrial Magnetism, Ionosphere  and Radio Wave Propagation, Russian Academy of
Sciences,  Troitsk Moscow Region 142190, Russia.}

\author{M.L. Khodachenko}
\address{Space Research Institute, Austrian Academy of Sciences, Graz 8042, Austria.}

\author{Wahab Uddin}
\address{Aryabhatta Research Institute of Observational Sciences (ARIES), Manora Peak,
Nainital-263 129, India.}

\begin{abstract}
We study the active region NOAA 10960, which produces two flare
events (B5.0, M8.9) on 04 June 2007. We find the observational
signature of right handed helical twists in the loop system
associated with this active region. The first B5.0 flare starts with
the activation of helical twist showing $\sim$3 turns. However,
after $\sim$20 minutes another helical twist (with $\sim$2 turns)
appears, which triggers M8.9 flare. Both helical structures were
closely associated with a small positive polarity sunspot in the AR.
We interpret these observations as evidence of kink instability,
which triggers the recurrent solar flares.
\end{abstract}

\section{Introduction}
Solar flares are transient explosions in the solar atmosphere when
the energy of stressed and twisted magnetic fields is released into
heating and radiation. The flares associated with coronal mass
ejections (CME) are known as ``eruptive flares", while flares
without CMEs are known as ``confined flares". The large flares
accompanied with energetic CMEs may be triggered by flux-rope
eruption with significant changes in the photospheric fields [\refcite{gary2004,liu2003}]. The
flares can also be initiated due to the filament interactions
followed by halo CMEs [\refcite{kumar2010a} and
references cited there] and filament eruptions [\refcite{liu2008} and references cited there]. Observations show
that the moderate flares without CME may be triggered by some
instabilities (e.g., kink instability) [\refcite{sri2010,kumar2010b}]. The
instabilities may cause the destabilization of large-scale magnetic
field, and can result CMEs [\refcite{cho2009} and
references cited there]. Instability of twisted magnetic flux tubes
are well studied in theory and intense numerical simulations [\refcite{torok2004,torok2005,hay2007}].

Kumar et al. [\refcite{kumar2010b}] have recently presented a
detailed multi-wavelength observations of the M8.9/3B class solar
flare in the active region NOAA 10960 on 04 June 2007. They
concluded that the positive flux emergence, the penumbral filament
loss of the associated sunspot and the activation of the several
twisted flux ropes in and around the flare site can be key
candidates for the occurrence of this flare during 05:06 UT and
05:13 UT. The ``activation" implies motion/brightening in 
the flux rope which is generated by some instability. 
A small B5.0 class flare has also been observed in the
same active region during 04:40-04:51 UT, which seems to be a
precursor for the M8.9 flare [\refcite{sri2010}]. These two
recurrent flares have been occurred during the activation of
successive helical twist and kink unstable flux tubes from a
positive polarity sunspot of AR 10960 [\refcite{sri2010,kumar2010b}].

In this paper, we review the occurrence of recurrent solar flares
and associated multi-wavelength phenomena. Multi-wavelength
observations are described in section 2. The observational results
are presented in section 3. The discussion and conclusions are given
in the last section.

\begin{figure}
\centering
\psfig{file=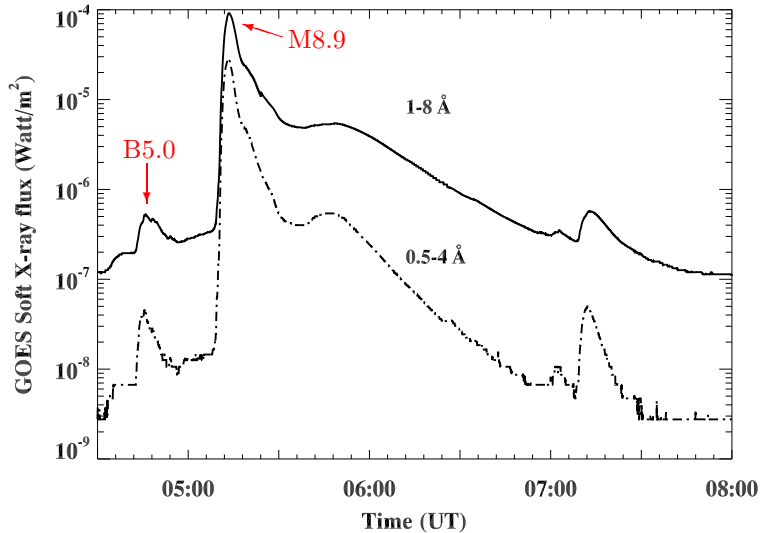,width=10cm}
$\color{red} \put(-235,140){\vector(0,-1){15}}\color{red} \put(-244,142){B5.0}$
$\color{red} \put(-187,188){\vector(-3,1){15}}\color{red} \put(-185,184){M8.9}$
\caption{GOES Soft X-ray flux profiles in two different wavelength bands for both flares (indicated by arrows) on 4 June 2007.}
\label{fig1}
\end{figure}

\begin{figure}
\centering
\psfig{file=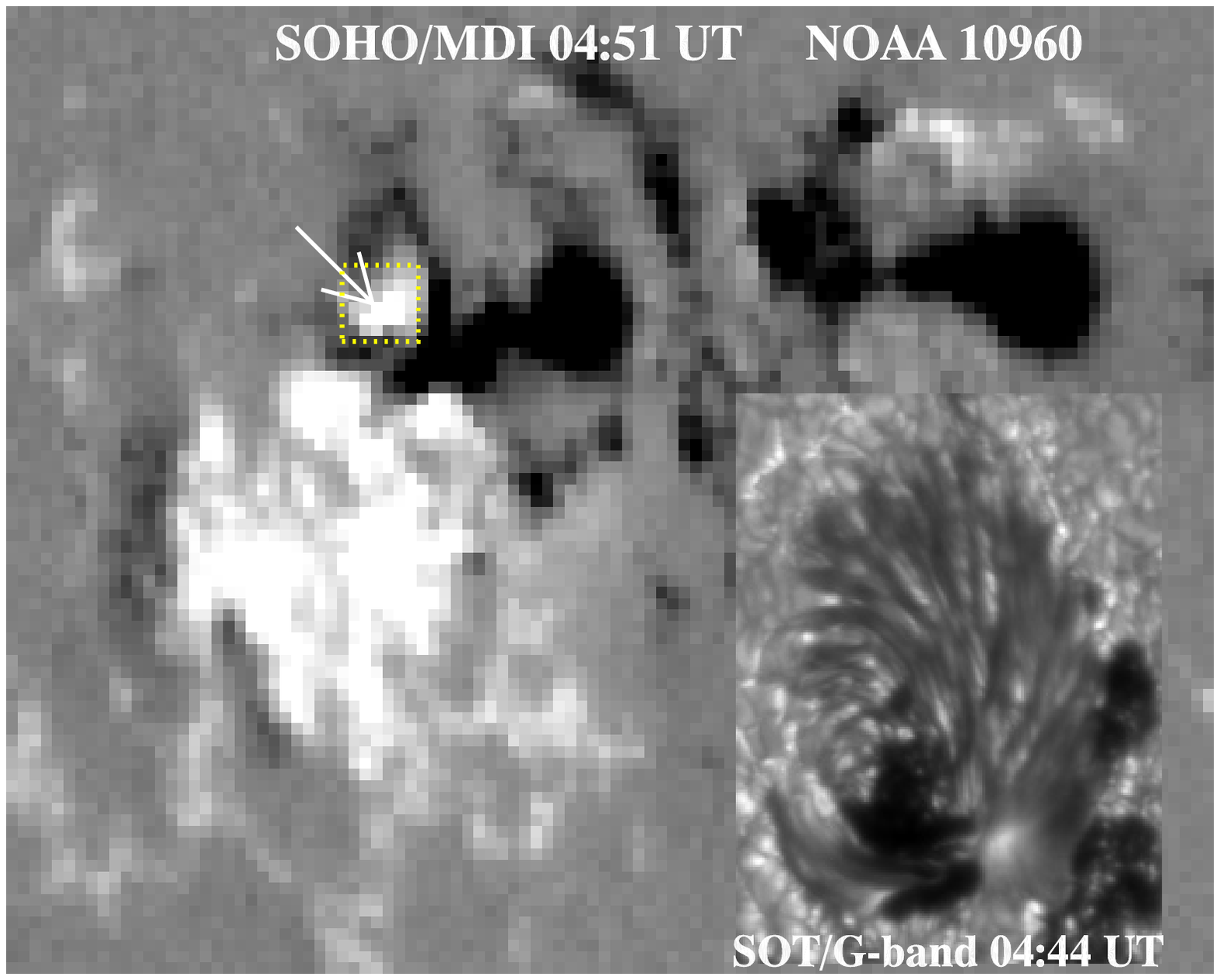,width=9cm}
\caption{SOHO/MDI magnetogram of the active region NOAA 10960 on 4 June 2007. The small positive polarity sunspot is shown in the box, indicated by an arrow. The enlarged view of the same sunspot is shown in Hinode SOT/G-band image in the bottom-right corner.}
\label{fig2}
\end{figure}

\section{Observational Data Sets}
{\bf SOHO MDI data:}  We use the SOHO/MDI data in order to show the
magnetic structure of the active region [\refcite{sch1995}]. The cadence of images is
96 minute and the pixel resolution is 1.98$^{\prime\prime}$.\\
{\bf TRACE data:} We use TRACE 171 \AA \ (Fe IX) EUV images to study the dynamics of the flaring active region
and its response in the corona during the flare event [\refcite{handy1999}]. This wavelength corresponds
to 1.3 MK plasma. The image size is 1024$\times$1024 pixels with resolution of 0.5$^{\prime\prime}$ per pixel,
and the cadence is $\sim$one minute. We have used the standard IDL routines available in the
SolarSoft library for cleaning and co-aligning the images.\\
{\bf Hinode SOT and XRT data:} We use Hindoe SOT G-band and Ca II H line data to
study the photospheric and chromospheric responses with an angular resolution
of 0.25 arcseconds, or 175 km over the field of view of about
400′$^{\prime\prime}$$\times$400$^{\prime\prime}$ on the Sun [\refcite{tsun2008}].
 For coronal study of the event, we use Hinode XRT data with an angular resolution of about 2$^{\prime\prime}$
 over a broad temperature range of 1--30 MK with full disk and a temporal resolution as short
as 2 s [\refcite{golub2007}].\\
{\bf STEREO SECCHI data:} For the impulsive phase of the flare, we use
the STEREO-A/SECCHI/EUVI observations [\refcite{wu2004}]. We use Fe IX 171 \AA \
coronal images for the present study. The size of each image is 2048$\times$2048 pixels with
1.6$^{\prime\prime}$ per pixel sampling. We use the standard SECCHI$\_$PREP subroutines for cleaning the
images and other standard subroutines available in STEREO package SolarSoft library.

\begin{figure}
{
\psfig{file=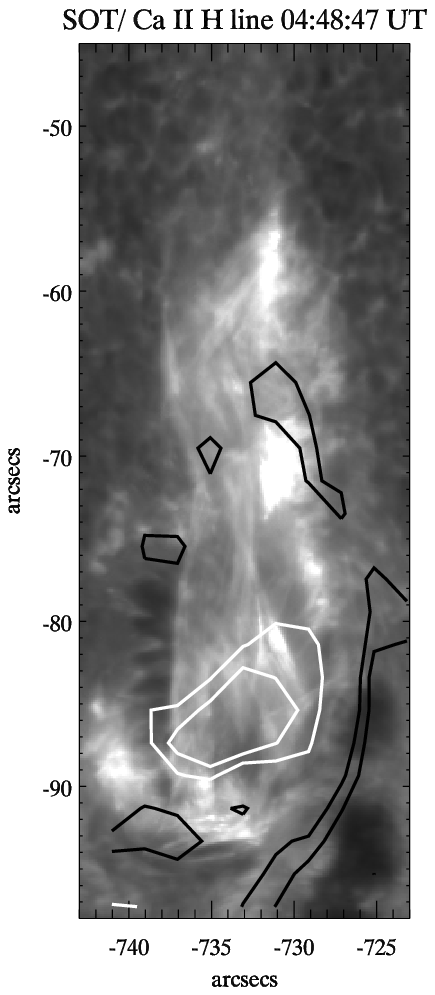,width=3.1cm}
$\color{red} \put(-65,-10){(a)}$
\psfig{file=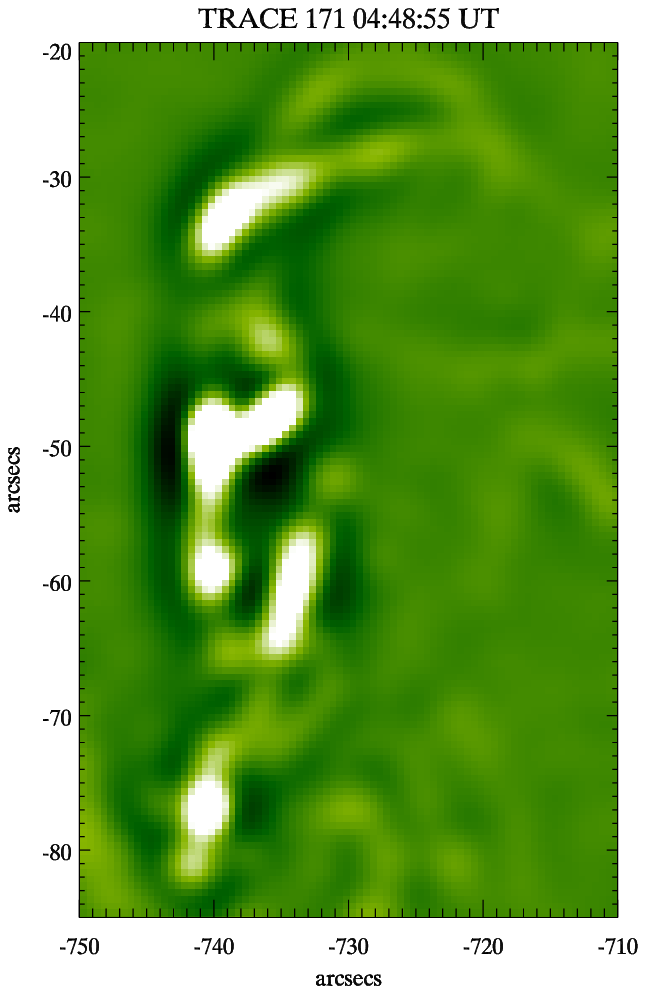,width=4.1cm}
$\color{yellow}\put(-35,80){\vector(-1,0){15}} \color{yellow}\put(-50,135){\vector(-1,0){15}}\color{yellow}\put(-50,35){\vector(-1,0){15}}\color{red} \put(-100,-10){(b)}$
\psfig{file=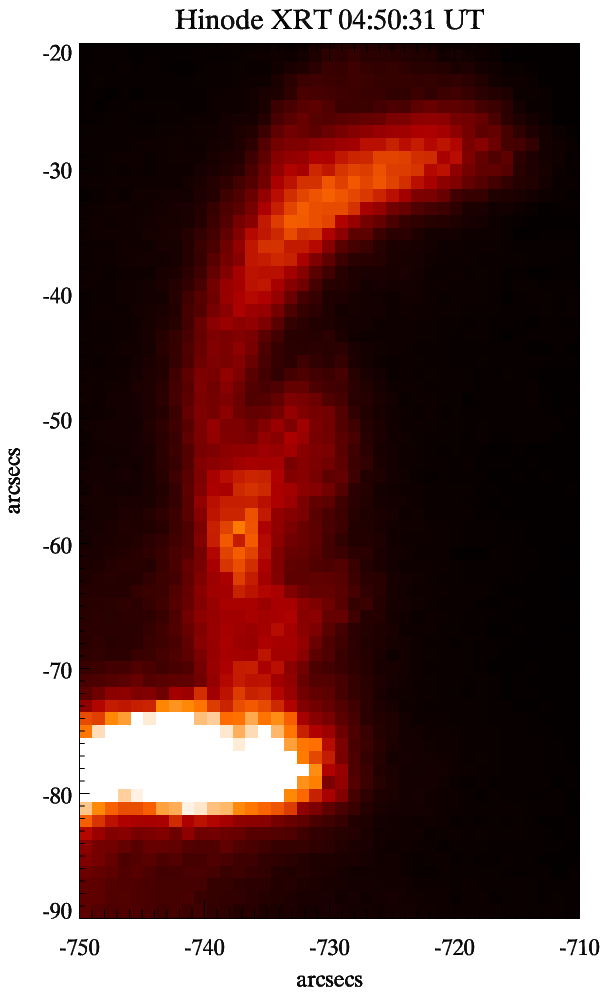,width=3.85cm}
$\color{red} \put(260,0){(c)}$
}
\caption{(a) Hinode SOT/Ca II H line image showing the chromospheric part of first helical twisted magnetic structure during the first B5.0 flare. SOHO MDI contours overlaid on this image shows that the twist activation is associated with small positive polarity sunspot (white=positive polarity, black=negative polarity). (b) TRACE 171 \AA \ EUV wavelet filtered image showing $\sim$3 turns (indicated by yellow arrows) in the helical twisted structure. (c) Hinode/XRT image showing the twisted structure during B5.0 flare.}
 \label{fig3}
\end{figure}

\begin{figure}
\centering{ \psfig{file=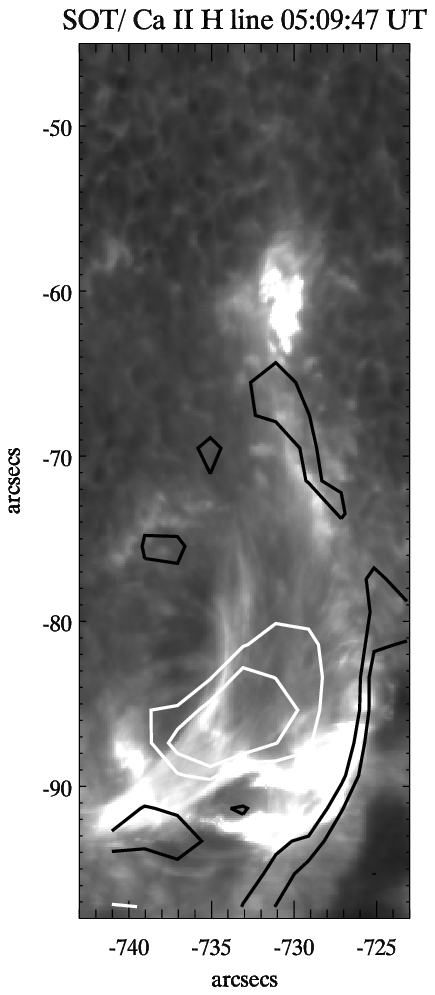,width=3.1cm} $\color{yellow}
\put(-65,165){(a)}$ \psfig{file=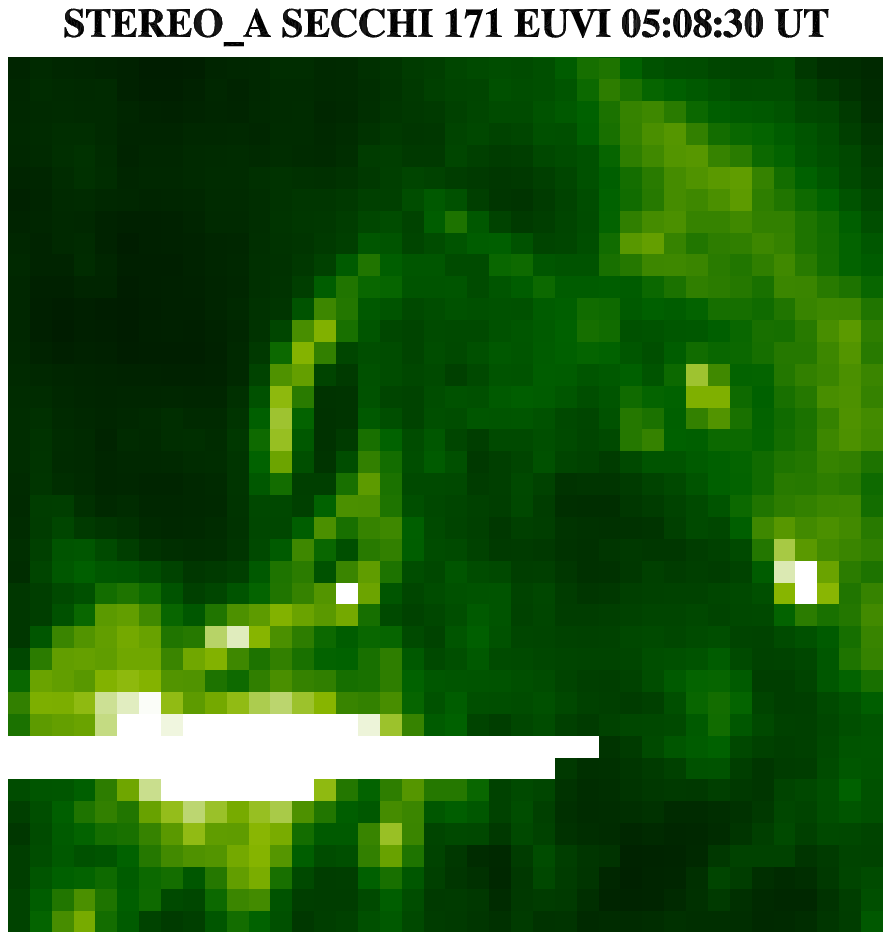,width=6.2cm}
$\color{yellow} \put(-100,165){(b)}$ \thicklines $\color{yellow}
\put(-120,140){\vector(0,-1){15}}\color{yellow}
\put(-170,150){Twisted helical structure}$ $\color{yellow}
\put(-85,80){\vector(-1,0){15}}$ $\color{yellow}
\put(-140,15){\vector(0,1){15}}\color{yellow} \put(-170,10){Flare}$
} \caption{(a) Hinode SOT/Ca II H line image showing the chromospheric 
part of second helical twisted magnetic structure during the second 3B/M8.9 flare. SOHO MDI
contours overlaid on this image shows that the twist activation is
associated with small positive polarity sunspot (white=positive
polarity, black=negative polarity). (b)(b) STEREO A SECCHI 171 \AA \ EUV image showing $\sim$2 turns in the helical twisted structure during M8.9 flare. The size of this image is 40$\times$40$^{\prime\prime}$.} 
 \label{fig4}
\end{figure}
\section{Observational Results}
Figure 1 displays the GOES Soft X-ray flux profile in two wavelength bands (1--8 and 0.5--4 \AA).
 Both flares (B5.0 and M8.9) are indicated by arrows. According to this profile, first B5.0 flare starts at 04:40 UT
, attains maximum at 04:43 UT and ended at 04:51 UT. Another M8.9
flare starts at 05:06 UT, maximizes at 05:13 and ended at
$\sim$05:30 UT. In H$\alpha$ classification, this flare was long
duration event classified as 3B, which starts at 05:05 UT and ends
at 06:42 UT. Figure 2 shows the complex active region (AR) NOAA
10960 in SOHO MDI image with $\beta\gamma\delta$ magnetic
configuration. This AR was very poor in CME production  and only two
M-class flares were associated with CMEs [\refcite{yashiro2008}]. The small positive polarity sunspot shown
in the box, indicated by an arrow plays the crucial role in
triggering recurrent flare activities in the AR. The enlarged view of the 
same sunspot is shown in Hinode SOT/G-band image in the bottom-right corner. Twisted helical
structures were activated above this sunspot during the initiation
of both flares. The enlarged view of this sunspot is shown in
SOT/G-band image in the bottom-right corner of this image. The
sunspot shows highly twisted penumbral filaments oriented in the
counter-clockwise direction.

The left-most panel of Figure 3 shows the chromospheric 
part of first helical twisted magnetic structure above the positive 
polarity sunspot during the B5.0
class solar flare at 04:48 UT. The MDI/SOHO contours are
over-plotted on the Ca II H 3968 \AA\ Hinode/SOT image. The white
contours show the positive magnetic polarity, while the black
contours show the negative magnetic polarity. The middle panel shows
TRACE 171 \AA \ EUV wavelet filtered image on the same time, which
shows the $\sim$ 3 turns of the helical twist above the same
positive polarity sunspot in the corona. Therefore, the total twist
angle, $\Phi\sim$6.0$\pi$ is much larger than the Kruskal -Shafranov
instability criterion ($\Phi$$>$$2.5 \pi$) for the generation of the
kink instability. The XRT images co-aligned with the TRACE, is
presented in the right-most panel of Figure 3. Sudden brightening
and enhancement of soft X-ray flux in the loop has also been evident
around 04:49-04:50 UT by XRT/Hinode. The loop is twisted during the
full span of the B5.0 flare duration, however, the twist maximizes
with the flare around $\sim$04:48--04:49 UT. The right-handed
(positive) twist is also clearly visible at 04:50 UT in the XRT
temporal image, which confirms the nature of twist as visible in the
EUV 171 \AA \ image from TRACE. The double structure of loop top in
171 \AA \ is also visible during the flare event.

\begin{figure}
\centering
\psfig{file=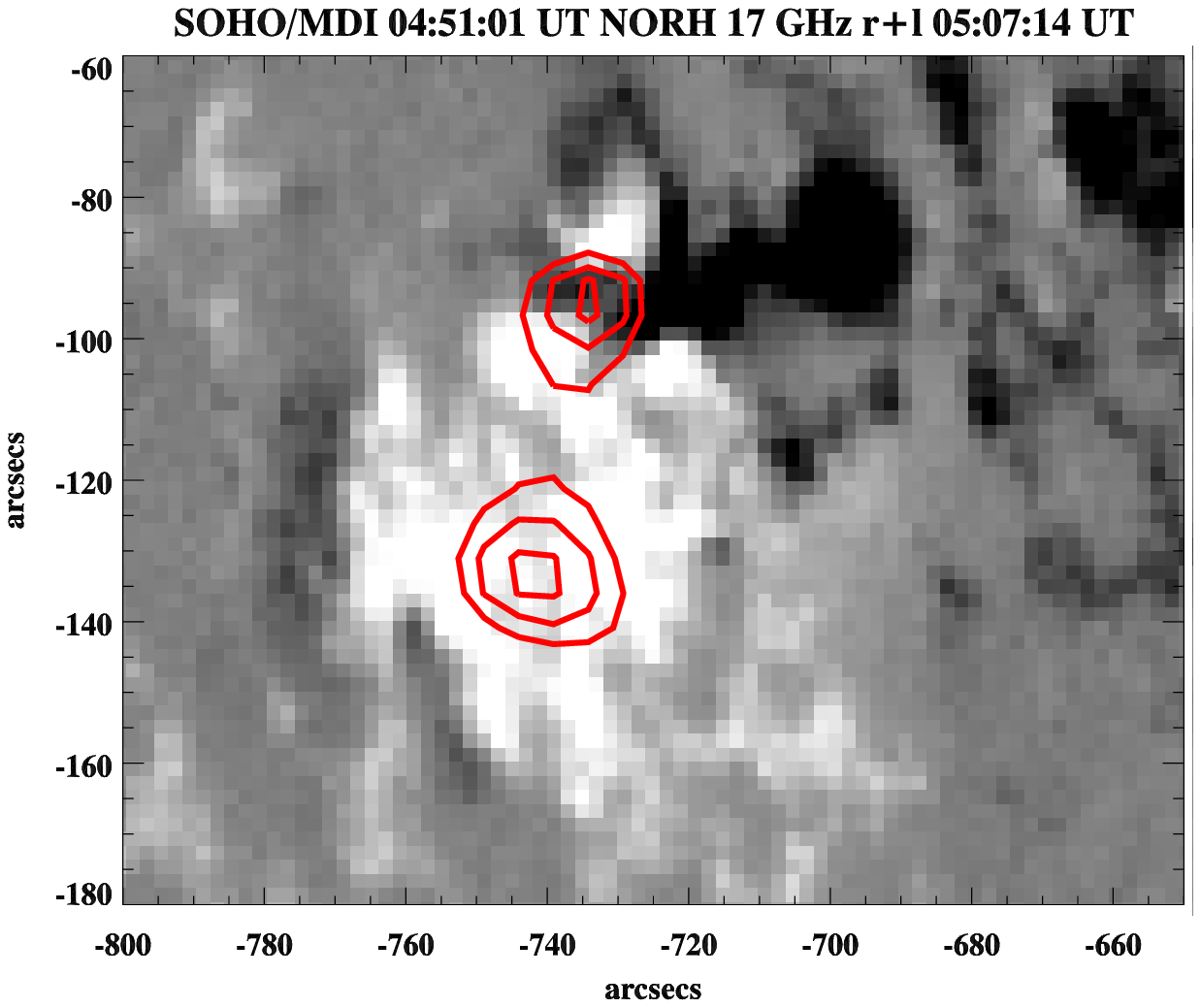,width=8cm}
\psfig{file=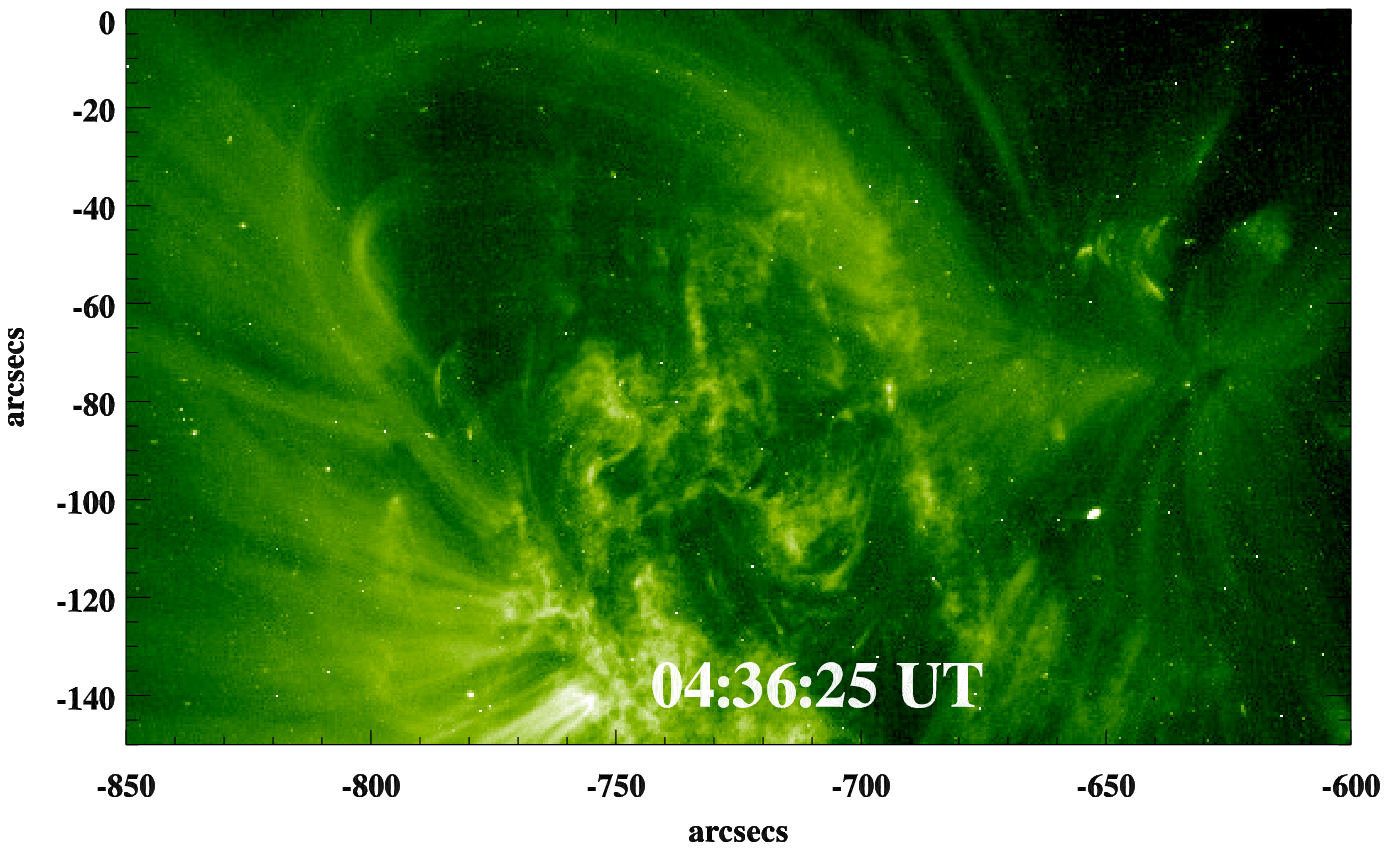,width=5.5cm}
\psfig{file=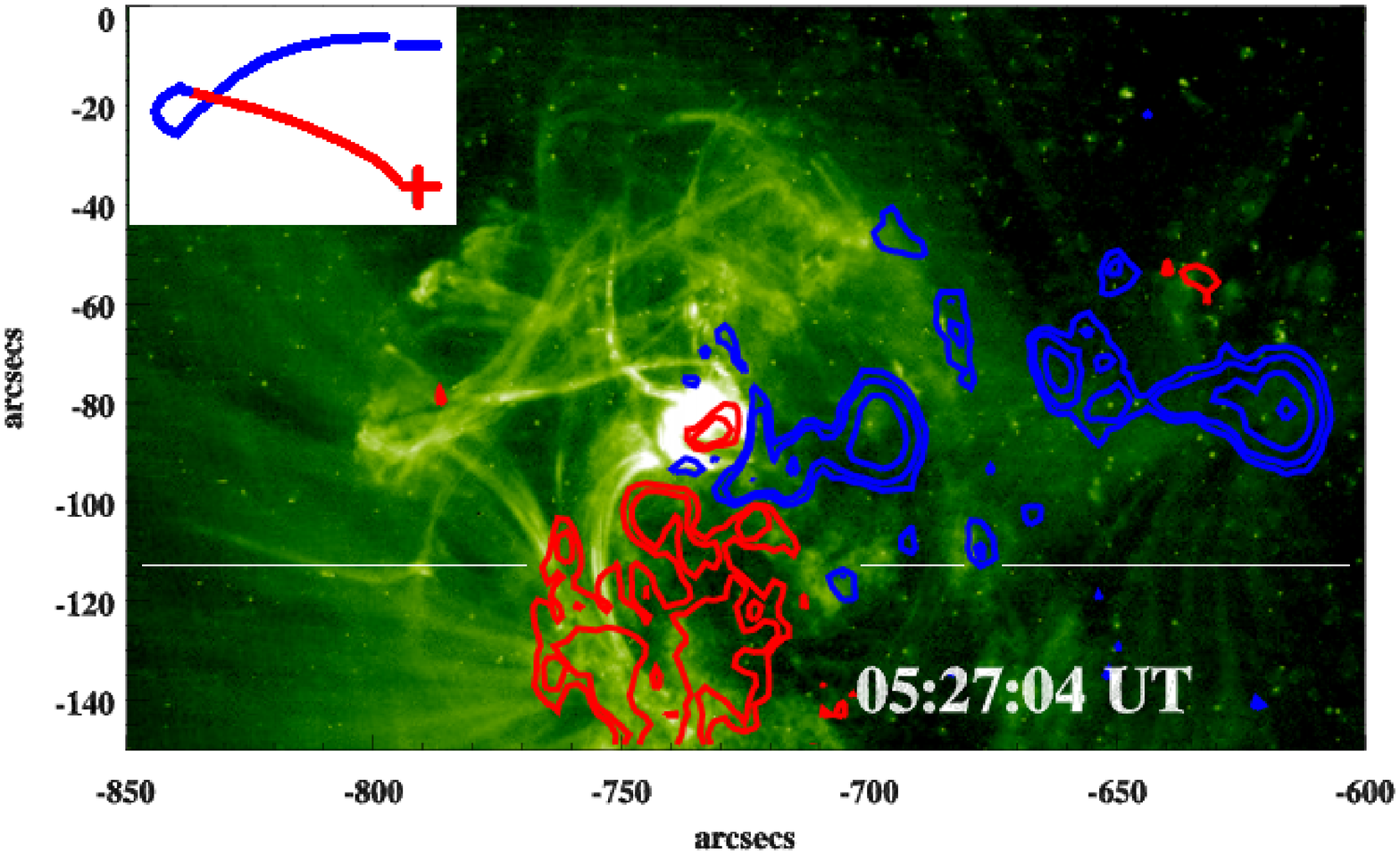,width=5.4cm} \caption{Top: Nobeyama
Radioheliograph (NoRH) 17 GHz contours overlaid on MDI image on 4
June 2007. Bottom: TRACE 171 \AA \ EUV images showing the magnetic
field environment in the AR before the flare (left panel) and during
the decay phase of the flare (right panel).} \label{fig5}
\end{figure}

The left panel of Figure 4 shows the chromospheric 
part of second helical twisted magnetic structure 
 above the same positive polarity sunspot during the
M8.9/3B class solar flare at 05:09 UT. The MDI/SOHO contours are
over-plotted on the Ca II H 3968 \AA\ Hinode/SOT image. The white
contours show the positive magnetic polarity, while the black
contours show the negative magnetic polarity. It is clearly evident
that M-class flare initiated above the same positive polarity
sunspot, where the B-class flare occurred. The right panel shows
STEREO-A/SECCHI EUVI 171 \AA \ image, which shows nearly $\sim$2
turns indicated by arrows in the second activation of the helical
twist on the same time above the positive polarity sunspot in the
corona. Therefore, the total twist angle, $\Phi\sim$4.0$\pi$ is
still much larger than the Kruskal -Shafranov instability criterion.
On the other hand, TRACE has no observations during the impulsive 
phase of the M8.9/3B flare.

Figure 5 (Top panel) shows the Nobeyama Radiheliograph (NoRH) 17 GHz
contours overlaid on the SOHO/MDI image. It shows the two radio
sources. One radio source is located close to small positive
polarity sunspot where the activation of helical twists takes place,
whereas another source is located at the center of major positive
polarity sunspot region. Bottom panels of Figure 5 shows the TRACE
171 \AA \ EUV images before the recurrent flares (04:36 UT, left
panel) and during the decay phase of the M-class flare (05:27 UT,
right panel). The left image shows a well organized loop-system in
the AR connecting opposite polarity regions. The right image shows
the SOHO/MDI contours overlaid on TRACE 171 \AA \ image
Red=positive, Blue=negative). The careful study of this image
reveals the activation of flux rope above the small positive
polarity sunspot, which is evident in the earlier high resolution
observations of this event (refer to Figures 3 and 4). The right
footpoint of the flux rope seems to be anchored in the negative
polarity field region. The knot (kink) structure and the morphology
of the flux rope is shown in the top-left corner of this image.

\section{Discussion and Conclusions}

We study the recurrent flare activities in the active region NOAA AR
10960, which are accompanied by activations of helical twists over
the small positive polarity $\delta$ sunspot.

Srivastava et al. [\refcite{sri2010}] have observed the first activation of a highly (right-handed)
twisted flux tube in AR 10960 during the period 04:43--04:52 UT. They have estimated
the length and the radius of the loop as L$\sim$80 Mm and a$\sim$4.0 Mm respectively, and also
estimated the total maximum twist angle as $\sim$12$\pi$ , by assuming quasi-symmetric distribution
 of the twist along the magnetic loop, which is much larger than the Kruskal--Shafranov
instability criterion. They suggested that the right-handed twist is
symmetrically distributed along the observed loop as a possible
asymmetry can be smoothed over the short Alfv\'en time of $\sim$80
s. The detection of a clear double structure of the loop top during
04:47--04:51 UT in TRACE 171 \AA \ images are found to be consistent
with simulated kink instability in curved coronal loops
(T{\"o}r{\"o}k et al. [\refcite{torok2004}]. They have suggested
that the kink instability of this twisted magnetic loop triggered
the B5.0 class solar flare, which occurred during 04:40 --04:51 UT
in this active region. The co-spatial brightening in soft X-rays as
observed by Hinode/XRT and the co-temporal occurrence of the
right-handed twisting in the flux tube confirm the occurrence of the
B5.0 flare during 04:40--04:51 UT probably due to the generation of
the kink instability.

Kumar et al. [\refcite{kumar2010b}] have found multi-wavelength
evidence of the successive activation of helical twists that may
help in the energy build-up process at the flaring region in
AR10960. The energy is released in the form of M-class flare after
secondary activation of helical twist in the flux tube when it
reconnects with neighboring opposite field. The activation of two
helical structures/ropes played an important role in destabilizing
and consecutive reconnection of magnetic field. The twist in the
secondary magnetic structure crosses the threshold (2.5--3.5$\pi$),
which probably produces the kink instability in this structure. The
energy release region in the M-class flare coincides with the
twisted magnetic structures. The M-class flare showed agreement with
the quadrupolar (closed--closed) reconnection model (breakout)
between two closed field lines [\refcite{anti1998}],
which is evident in the decay phase of this flare (see bottom-right
panel of Figure 5 also). The asymmetric evolution is driven by
foot-point shearing of one side of an arcade, where reconnection
between the sheared arcade and the neighboring (unsheared) flux
system most probably triggers the flare. The kink instable twisted 
magnetic structure may undergo in a weak reconnection with the surrounding closed 
field lines in quadrupolar field configuration. Therefore, it triggers M-class flare in the active region without
 any eruption/CME [\refcite{anti1999,aul2000,asc2004}].

Disappearance of Penumbrae
during the decay phase and after the flare suggests that the
magnetic field changes from inclined to almost vertical
configuration [\refcite{kumar2010b}]. This means
that the part of penumbral magnetic field is converted into umbral
fields. These results are in agreement with previous studies [\refcite{wang2004,liu2005}]. The
rotation of sunspot was the most plausible cause of the helical
twist and thus energy release in B5.0 class flare [\refcite{sri2010}]. On the other hand, the penumbral loss and
umbral area enhancement were clearly evident during M-class flare
[\refcite{kumar2010b}].

Nobeyama 17 GHz radio contours overlaid the SoHO/MDI image during
the M-class flare show the two radio sources, corresponding to the
footpoints of magnetic loop system, which are generated during the
impulsive phase of the M-class flare due to particle acceleration from
the reconnection site. The bottom-right panel of Figure 5 shows
the TRACE 171 \AA\ image in which the same bright loop system is
clearly evident in the southward direction. The part of rising helical 
structure most likely reconnects with the 
 southward loop-system and produce two radio sources due to particle 
acceleration from the reconnection site. These radio sources are the footpoints 
of the flaring loop system. The existence of co-spatial radio sources with this
loop-system suggests the reconnection of twisted flux rope with 
the ambient field as most possible scenario for
the flare triggering [\refcite{kumar2010b}].

In conclusions, this paper reviews the rare observational signature of 
kink instability associated with failed eruptions and solar flares. 
Earlier, several researchers have been reported kink instability 
associated with CME eruptions [\refcite{liu2008,cho2009} and references cited there].
 The very interesting active region AR10960 was poor
CME generator, but triggered many solar flares during its journey
over the solar-disk. The multi-wavelength signature of helically
twisted structures has been found as the cause of the recurrent
solar flares on 04 June 2007. Therefore, we suggest that such flares
may occur due to some instability/activation of twisted magnetic
fields. The detailed multi-wavelength and statistical studies should
be performed in future with the high- resolution space borne and
ground-based observations.

\section *{Acknowledgments}
We thank to the reviewers for their constructive suggestions. 
We are thankful for the data from different space-based instruments
i.e. SOHO, GOES. SOHO is a project of international cooperation
between ESA and NASA. The radio data from Nobeyama is thankfully acknowledged.
This work was supported by the Department of
Science and Technology, Ministry of Science and Technology of India,
by the Russian foundation for Basic Research (grants 09-02-00080 and
09-02-92626, INT/RFBR/P-38) and by the Austrian Fond zur F\"orderung
der Wissenschaftlichen Forschung (project P21197-N16). The work of
T.Z was also supported by the Georgian National Science Foundation
grant GNSF/ST09/4-310.


\bibliographystyle{ws-procs9x6}
\bibliography{reference}

\end{document}